\documentclass[12pt,twoside]{article}
\usepackage{cite}
\usepackage{graphics}

\def\refj#1{{\it #1}}

\def\refjournal#1#2#3#4{{\it #1} {\bf #2} (#3) #4}

\begin{document}

\newcommand{\Lslash}[1]{ \parbox[b]{1em}{$#1$} \hspace{-0.8em}
                         \parbox[b]{0.8em}{ \raisebox{0.2ex}{$/$} }    }
\newcommand{\Sslash}[1]{ \parbox[b]{0.6em}{$#1$} \hspace{-0.55em}
             1
              \parbox[b]{0.55em}{ \raisebox{-0.2ex}{$/$} }    }
\newcommand{\Mbf}[1]{ \parbox[b]{1em}{\boldmath $#1$} }
\newcommand{\mbf}[1]{ \parbox[b]{0.6em}{\boldmath $#1$} }
\newcommand{\beq}{\begin{equation}}
\newcommand{\eeq}{\end{equation}}
\newcommand{\beqa}{\begin{eqnarray}}
\newcommand{\eeqa}{\end{eqnarray}}
\newcommand{\skipfields}{\!\!\!\!\! & \!\!\!\!\! &}

\newcommand{\gsim}{\buildrel > \over {_\sim}}
\newcommand{\lsim}{\buildrel < \over {_\sim}}
\newcommand{\ie}{{\it i.e.}}
\newcommand{\eg}{{\it e.g.}}
\newcommand{\cf}{{\it cf.}}
\newcommand{\etal}{{\it et al.}}
\newcommand{\gev}{{\rm GeV}}
\newcommand{\jpsi}{J/\psi}
\newcommand{\order}[1]{${\cal O}(#1)$}
\newcommand{\eq}[1]{eq.~(\ref{#1})}

\newcommand{\tr}{T}
\newcommand{\ptr}{p_\tr}
\newcommand{\as}{\alpha_s}

\newcommand{\RS}{R_S(0)}
\newcommand{\Pqg}{P_{{\!}_{q\to g}}}
\newcommand{\psq}{\vec{p}_\perp^{\; 2}}
\newcommand{\subjet}{{\mbox{\scriptsize jet}}}
%
\newcommand{\LevelOne}[1]{ \subsection*{#1} }
\newcommand{\LevelTwo}[1]{ \subsubsection*{#1} }
\newcommand{\LevelThree}[1]{ \paragraph*{#1} }
\newcommand{\AppendixLevel}[2]{
   \subsection*{Appendix #1 -- #2}
   \markright{Appendix #1}
   \addcontentsline{toc}{subsection}{\protect\numberline{#1}{#2}} }
\def\todo#1{$\rightarrow$ {\bf #1}}
\renewcommand{\textfraction}{0.1}
\renewcommand{\floatpagefraction}{0.9}

\begin{titlepage}
\begin{flushright}
        NORDITA-96/40 P
\end{flushright}
\begin{flushright}
        hep-ph/9606472 \\ June 28, 1996
\end{flushright}
\vskip .8cm
\begin{center}
{\Large Generating heavy quarkonia\\
  in a perturbative QCD cascade}
\vskip .8cm
{\bf Per Ernstr\"om and Leif L\"onnblad}
\vskip .5cm
NORDITA\\
Blegdamsvej 17, DK-2100 Copenhagen \O\\
{\small per@nordita.dk, leif@nordita.dk}
\vskip 1.8cm
\end{center}
\begin{abstract}
\noindent
    We present an implementation of heavy quarkonium production within
    a perturbative QCD cascade based on the Color Dipole Cascade
    model.
    We consider the processes most relevant in the context of the
    $\psi'$ surplus at the Tevatron;
    $g\rightarrow\psi'$ and $c\rightarrow\psi'$ in the
    color-singlet model and
    $g\rightarrow\psi'$ through the color-octet mechanism.
    Our implementation is, however, easily extendible
    to other quarkonia and other production mechanisms.

    Where comparison is possible we find good agreement with analytical
    calculations.

    We present some suggestions for measurements at the Tevatron that
    would be sensitive to the shape of the fragmentation functions.
    Our calculations indicate that such measurements could be used
    to test the color-octet mechanism solution to the $\psi'$ surplus
    at the Tevatron.
\end{abstract}
\end{titlepage}

\setlength{\baselineskip}{7mm}
\raggedbottom

\section{Introduction}

Recent Tevatron data \cite{CDF,D0} on prompt charmonium
($\psi$ and $\psi'$)
production at large $\ptr$ has attracted a lot of attention.

At large $\ptr$, charmonium production is believed to be dominated by
parton fragmentation \cite{BraatenYuan}, and the Tevatron data
\cite{CDF,D0} does indeed show a $1/\ptr^5$ behavior consistent with
the predictions of a fragmentation process.

The large c quark mass $m_c$ make, however, predictions possible also
for the magnitude of the cross section. Production of charmonium
states can be factorized into the hard QCD production of a heavy
$c\bar{c}$ pair within a region $1/m_c$, followed by the
nonperturbative formation of the charmonium state. In \cite{BBL}, a
general factorization scheme was developed to all orders in the
relative velocity $v$ of the $c$ and $\bar{c}$ quark in the charmonium
state. For S-wave states, treated to leading order in $v^2$, this
factorization scheme reduces to the color singlet model \cite{Baier};
the $c\bar{c}$ pair must be produced in a color singlet state with
the same spin and angular momentum as the charmonium state. The
subsequent nonperturbative formation of the charmonium state can be
described by a single process independent parameter. Fixing this
parameter from measurements on the electro-magnetic decay of the $\psi$
and $\psi'$, predictions can be made.

A color singlet model calculation to LO in $\as$ of large $\ptr$
$\psi'$ production fall, however, more than an order of magnitude
below the Tevatron data
\cite{CDF,D0,BDFMhighptSingletTheory,RoySridhar}. Although
this is the largest deviation from LO color singlet model predictions
seen, it should be noted that large deviations have been observed also
at low $\ptr$ \cite{Schuler,VHBT}.

A calculation of prompt $\psi$
production \cite{BDFMhighptSingletTheory,RoySridhar,CacciariGreco}
give a similar result, but
in this case contributions from P-wave charmonium states decaying to
$\psi$ obscure the interpretation of the data.

Braaten \etal ~\cite{BraatenFleming} have proposed that the $\psi'$
surplus at the Tevatron could be explained by a $v$ suppressed
production mechanism, usually referred to as the color octet
mechanism. The $c\bar{c}$ pair is produced in a color octet $^3S_1$
state, and forms the $\psi'$ through soft gluon emission or
absorption. The new nonperturbative parameter giving the
``probability'' for the color octet $c\bar{c}$ pair to form the
$\psi'$ is supressed by $v^4$ as compared to the color singlet
mechanism. The hard gluon fragmentation production of a color octet
$c\bar{c}$ pair is, on the other hand, of lower order in $\as$ as
compared to the production of a $^3S_1$ color singlet $c\bar{c}$
pair.

In addition, the color octet fragmentation process is enhanced by the
so called trigger bias effect. The fragmentation contribution to the
charmonium production cross section can be written as a convolution
\beq
  {d\sigma_{\cal O}(p) }
  = \sum_i
    \int_0^1 dz
    {d\sigma_i }(p/z,\mu)
    D_{i \rightarrow {\cal O}}(z,\mu),
\label{sigmap}
\eeq
of parton production cross sections ${d\sigma_i }$ and fragmentation
functions $D_{i \rightarrow {\cal O}}$, describing the fragmentation
of a parton $i$ into a charmonium state $\cal O$. Since the parton
production cross sections fall steeply with the parton transverse
momentum $\ptr/z$, fragmentation processes where the charmonium state
takes a large fraction $z$ of the parton momentum are favored.
In a strict LO calculation the color octet fragmentation function is
proportional to $\delta(1-z)$. When leading logs are re-summed,
fragmentation functions are softened, but the fact remains that in
comparison with other production mechanisms the color octet mechanism
is characterized by large $z$ fragmentation.

The arguments in favor of the color octet solution are of a
qualitative nature, given a new nonperturbative parameter one can
obviously fit to data. It is therefore crucial to find independent
means to check the hypotheses by Braaten \etal.

One way would be to look for other processes where the same color
octet production mechanism could be important, see eg.~\cite{epem}.
Alternatively one might look for characteristics of the color octet
production mechanism at the Tevatron. It has eg.\ been noted that
$\psi'$ produced through the color octet mechanism are dominantly
transversely polarized \cite{polarization}.

Here we will make use of the fact that a $\psi'$ produced in a
fragmentation process will be part of a jet. We investigate the
possibilities to extract more information on the fragmentation process
by detecting the associated jet. At a superficial level one can
identify the jet momentum with the parton momentum and thus get a
measure of the fraction $z$ of the parton momentum taken by the
$\psi'$. As noted above, large $z$ fragmentation is a characteristic of
the color octet production mechanism.

In \cite{HigherOrder} it was proposed that trigger bias enhanced higher
order corrections might explain the $\psi'$ surplus within the
color singlet model.
It was shown explicitly that trigger bias enhanced higher
order corrections to light quark fragmentation into $\eta_c$
dominate over the LO result.
For $\psi$ and $\psi'$ higher order calculations are however lacking.
If this proposal is correct
one would expect the fragmentation to be intermediate in hardness
between the soft LO color singlet mechanisms and the very hard
color octet mechanism.

In order to make a quantitative analysis it is necessary to implement
the fragmentation processes in a perturbative QCD cascade,
allowing the use of measurable variables such as
$z_{\subjet}=p_{\tr \psi'}/p_{\tr\subjet}$.

In this paper we present an implementation of the LO color octet and
singlet production mechanisms relevant for $\psi'$ production, but the
extension to other mechanisms and other quarkonia is trivial.

In section \ref{sec:implementation} we describe the implementation of
fragmentation production of quarkonium states in a leading--log QCD
shower generator based on the Dipole Cascade Model (DCM), first in
general and then more specificly for the LO color octet and singlet
production mechanisms relevant for $\psi'$ production.

In section \ref{sec:what} we give some results of our simulations
and discuss the possibilities to distinguish between different
fragmentation mechanisms based on the shape of fragmentation
functions.

Our conclusions are presented in section \ref{sec:conclusions}, which
is followed by an appendix describing the analytical approximations
used to verify our implementation.

\section{The Implementation}
\label{sec:implementation}

The fragmentation production of quarkonia may be viewed as a two step
process:
1. the {\em hard} fragmentation into a color singlet or
octet ${c\bar{c}}$-pair, and
2. the {\em nonperturbative} formation of the quarkonium state.

Here we implement both these steps in conjunction in a hard QCD
shower generator, by including splitting functions not
into ${c\bar{c}}$-pairs but directly into quarkonium states.
Thus, the nonperturbative formation of the quarkonium state is implemented
at the perturbative level, before normal hadronization sets in.
This is motivated by the fact that all hard interactions are
already included in the splitting functions.
The ${c\bar{c}}$-pairs are thus effectively noninteracting until
the hadronization onset scale.
Finally, according to BBL factorization \cite{BBL}, the ``probability''
that a color octet or singlet ${c\bar{c}}$-pair form a quarkonium
state is given by process independent parameters (matrix elements).
BBL factorization is however fully inclusive and does not describe
how the ${c\bar{c}}$-pair effects the hadronization of
the remaining partons.
In the color singlet case, the ${c\bar{c}}$-pair is color disconnected
from the other partons in the event and it is certainly reasonable
to let the other partons hadronize independently of the ${c\bar{c}}$-pair.
The color octet case is more troublesome.
In our implementation we enforce the radiation of a soft gluon that
inherits the color connection of the ${c\bar{c}}$-pair to the
other partons.
That is, we simply implement splitting into the quarkonium state plus
a soft gluon.
The hadronization is done using the string\-fragmentation model in
{\sc Jetset} which is infrared safe w.r.t.\ soft gluons, suggesting
insensitivity to the exact modelling of these soft interactions.
We will, however, return to this nontrivial question in a forthcoming
publication.

In this paper we present
an implementation of heavy quarkonium production in the Ariadne
program \cite{Ariadne} which implements the Dipole Cascade Model (DCM)
\cite{CDMinit86,CDMplain88} for QCD cascades.

The DCM differs from conventional parton shower models in that gluon
emission is treated as dipole radiation from color dipoles between
partons. In eg.\ $e^+e^-$ annihilation, the first gluon, $g_1$, is
emitted from the dipole between the initial $q\bar{q}$ pair. An
emission of a second softer gluon can then be described in terms of
radiation from two independent dipoles, one between the $q$ and $g_1$
and one between $g_1$ and $\bar{q}$. Further gluon emissions are given
by three independent dipoles etc.

One problem with this model is that only gluon radiation is modelled
and the process of splitting a gluon into a $q\bar{q}$ pair is not
included in a natural way. In \cite{CDMsplit90} it was shown that this
process can be included without leaving the dipole picture by taking
the ordinary splitting kernel for $g\rightarrow q\bar{q}$ and dividing
it in two equal parts describing the contribution from each connecting
dipole to the splitting separately. This procedure works very well and
describes correctly the amount of secondary heavy quarks produced at
LEP \cite{OPALg2cc95} although some theoretical questions have been
raised \cite{Mikeg2qq95}.

The process of $g\rightarrow {\cal O}+g$ will here be treated in the
same way as the $g\rightarrow q\bar{q}$ one. The splitting kernel
$d_{g\rightarrow {\cal O}}(z,S)$, which is related to the fragmentation
function through
\begin{equation}
  D_{g\rightarrow {\cal O}}(z,\mu) = \int_{0}^{\mu^2}
  \frac{dS}{S} d_{g\rightarrow {\cal O}}(z,S)
\end{equation}
is hence divided into two equal parts assigned to each of the
connecting dipoles. Assuming a dipole between a quark and a gluon in
its center of mass system, with the gluon along the positive z--axis,
$z$ is then the fraction of the total positive light-cone momenta of
the dipole carried by the quarkonium.  The partons in the DCM is
always on mass shell, therefore, $S$ is no longer the mass square of
the split gluon, but is still the mass square of the decay products.
Four-momentum is conserved by ``borrowing'' momentum from the other
parton in the dipole -- this is a valid approach in the limit of
strongly ordered emissions.

For practical reasons, the generations of quarkonium production is
done in the variables $m_\perp^2 \equiv m_{\cal O}^2 + p_\perp^2 = zS$
and $y=\ln{\frac{zW^2}{S}}$, the transverse mass and rapidity of the
quarkonium in the dipoles center of mass, W is the total mass of the
emitting dipole.

For eg.\ a $q$--$g$ dipole there are now several competing processes
that can occur; a gluon can be emitted according to the standard DCM,
the initial gluon can be split into a $q\bar{q}$ pair of different
flavors and it can be split into a quarkonium plus gluons according
to several different mechanisms. This competition is as usual governed
by the Sudakov formfactor mechanism using the transverse momentum
squared as ordering variable. Hence the probability $P_i(k_\perp^2)$
of the process $i$ to occur at a scale $k_\perp^2$ is given by
\begin{equation}
  \label{eq:sudakov}
  P_i(k_\perp^2) =
  d_i(k_\perp^2)
  e^{-\int^{k_{\perp,\max}^2}_{k_{\perp}^2}\sum_jd_j(p_\perp^2)dp_\perp^2}
\end{equation}
where $d_i(k_\perp^2)$ is the splitting kernel with the dependence on
rapidity integrated out.

It is now straight forward to implement basically any mechanism for
fragmentation production of any heavy quarkonium. In this paper we
discuss three mechanisms for $\psi'$ production:
\begin{itemize}
\item Color-Octet mechanism $g\rightarrow c\bar{c} (^3S_1^{(8)})
  \rightarrow \psi' \pm$ soft gluons.
  A perturbative transition of a gluon into a
  collinear $c\bar{c}$ pair in a color octet $^3S_1$ state, followed
  by a nonperturbative transition into $\psi'$ involving emission
  and/or absorption of soft gluons.
  As discussed above we implement this process as $g\rightarrow \psi'+g$,
  with one soft final state gluon
{}.
  Since the gluons are soft, the splitting kernel is essentially a
  delta-function,
\beq
   d_{g\rightarrow {\psi'}}(z,S)=
   {\pi \alpha_s \langle0|O^{\psi'}_8(^3S_1)|0\rangle \over 8 m_c}
   \delta(1-z)\delta(S-4m_c^2).
\eeq
  In the implementation we give for practical reasons the splitting kernel
  a small width, but we do not in this paper study possible effects of
  the physical ${\cal O}(v^2)$ width \cite{BraatenRev}.
  Since the characteristic velocity $v$ of the c-quarks in the charmonium
  state is not that small ($v^2\approx0.2$),
  the physical width could effect the event structure and through trigger bias
  enhancement also the normalization of the large $\ptr$ quarkonium production.
  The Color-Octet matrix element $\langle0|O^{\psi'}_8(^3S_1)|0\rangle$
  was taken to be $0.0042$ GeV$^3$ as given by the fit in
  \cite{BraatenFleming} to CDF data on large $\ptr$ $\psi'$
  production.
\item Color-Singlet mechanism
  $g \rightarrow c\bar{c}(^3S_1^{(8)})+2g \rightarrow \psi'+2g$ is
  theoretically simpler since the hard splitting color disconnects
  the $c\bar{c}$-pair from the rest of the event.
  The implementation is complicated, however, due to the final 3-particle
  phase space.  For simplicity we treat also this process as
  $g\rightarrow \psi'+g$. The error introduced by this
  simplification, which will be investigated in a future publication,
  may be assumed to be small since two gluons nearby in phase space
  are usually well approximated with one within the framework of Lund
  string fragmentation. The color-singlet gluon fragmentation
  function is given as a double integral in \cite{BraatenYuanSwave}.
  Performing one of the integrals we find an analytical expression for
  the fictitious $g\rightarrow \psi'+g$ splitting kernel.  The color
  singlet parameter $|R(0)|^2$ relevant also for the last production
  mechanisms is taken to be $0.231$ GeV$^3$.
\item Color-Singlet mechanism for $c\rightarrow \psi' + c$. Here
  the whole splitting kernel, given in \cite{BraatenCheungYuan}, is
  assigned to the only dipole connected to the $c$ quark. The scale is
  the running $\alpha_S$ is, as in the other cases, taken to be
  $m_\perp^2$.
\end{itemize}

\section{What comes out}
\label{sec:what}

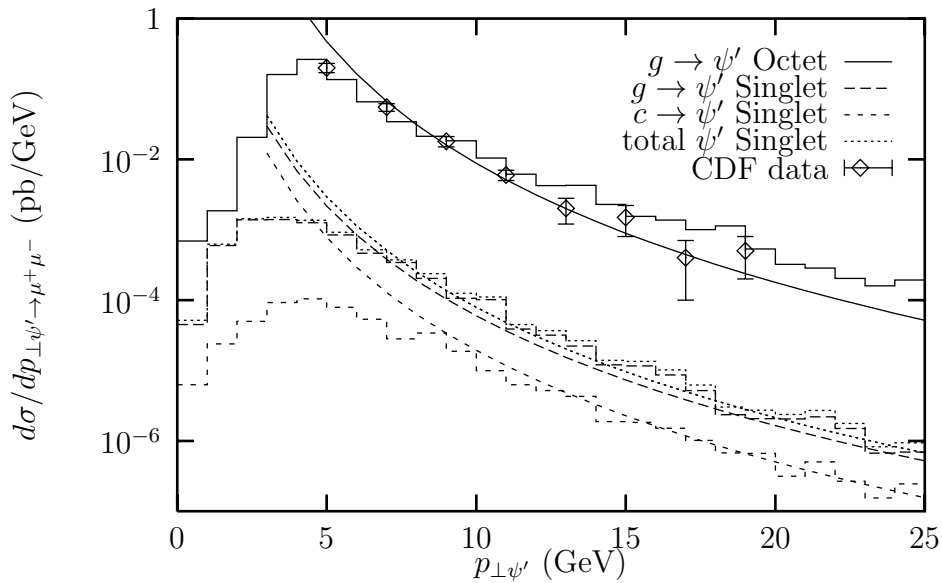
\begin{figure}

\setlength{\unitlength}{0.1bp}
\begin{picture}(3600,2160)(0,-100)
\put(3054,1546){\makebox(0,0)[r]{CDF data}}
\put(3054,1646){\makebox(0,0)[r]{total $\psi'$ Singlet}}
\put(3054,1746){\makebox(0,0)[r]{$c\rightarrow \psi'$ Singlet}}
\put(3054,1846){\makebox(0,0)[r]{$g\rightarrow \psi'$ Singlet}}
\put(3054,1946){\makebox(0,0)[r]{$g\rightarrow \psi'$ Octet}}
\put(2008,51){\makebox(0,0){$p_{\perp\psi'}$ (GeV)}}
\put(100,1180){%
\makebox(0,0)[b]{\shortstack{$d\sigma/dp_{\perp\psi'\rightarrow\mu^+\mu^-}$
(pb/GeV)}}%
}
\put(3417,151){\makebox(0,0){25}}
\put(2854,151){\makebox(0,0){20}}
\put(2290,151){\makebox(0,0){15}}
\put(1727,151){\makebox(0,0){10}}
\put(1163,151){\makebox(0,0){5}}
\put(600,151){\makebox(0,0){0}}
\put(540,516){\makebox(0,0)[r]{$10^{-6}$}}
\put(540,1047){\makebox(0,0)[r]{$10^{-4}$}}
\put(540,1578){\makebox(0,0)[r]{$10^{-2}$}}
\put(540,2109){\makebox(0,0)[r]{$1$}}
\end{picture}
\vskip -7mm
\caption[dummy]{{\it The transverse momentum distribution of
    $\psi'\rightarrow \mu^+\mu^-$ production at the Tevatron for
    different production mechanisms compared to CDF data
    \cite{CDF}. Full line is the color octet mechanism, dotted
    line is the sum of the color singlet mechanisms, and the
    long-dashed and short-dashed lines are the gluon and c-quark
    initiated singlet mechanisms respectively. The histograms were
    generated with {\sc Ariadne} and the smooth lines are from the
    analytical approximation in the Appendix.}}
\label{fig:ptspec}
\end{figure}

The {\sc Ariadne} program is interfaced to {\sc Pythia} and {\sc
  Jetset} \cite{JETPYT93,JETPYT94} for handling of the hard
sub-process and hadronization respectively. We use this to generate
$\psi'$ production events at the Tevatron. In fig.~\ref{fig:ptspec} we
see the resulting $p_{\tr \psi'}$ spectrum for the different
production mechanisms. Using the same parameters as in
\cite{BraatenFleming}, we reproduce approximately their result where
the Octet mechanism totally dominates the cross section for high
$p_{\tr\psi'}$, and we also reproduce the measured
$p_{\tr\psi'\rightarrow\mu^+\mu^-}$ spectrum \cite{CDF}.

The difference between the Octet and Singlet mechanisms is larger in
our case than in \cite{BDFMhighptSingletTheory,BraatenFleming}
even after taking into account the small variations in the parameters used.
The generated spectra from {\sc  Ariadne} does, however, agree
with the analytical approximation in
the appendix at large $p_{\tr}$. At small $p_{\tr}$ we do not expect
these to agree because there the simple assumption in the analytical
calculation that the gluon spectrum behaving like $1/p_{\tr}^5$
breaks down. Also effects of initial-state QCD radiation, implemented
through the so-called Soft Radiation Model \cite{CDMdis89} in {\sc
  Ariadne}, are absent in the analytical calculation.
In addition, {\sc Ariadne} implements a more complete treatment of the parton
evolution, in contrast to the analytical calculations, which, like
earlier numerical calculations, uses the approximative though
asymptotically correct AP evolution.
The full evolution equation and a discussion of the approximation made
in the use of the AP evolution equation can be found
in \cite{BDFMhighptSingletTheory}.

\begin{figure}
\setlength{\unitlength}{0.1bp}
\begin{picture}(3600,2160)(0,-100)
\put(1727,1496){\makebox(0,0)[r]{$c\rightarrow \psi'$ Singlet}}
\put(1727,1596){\makebox(0,0)[r]{$g\rightarrow \psi'$ Singlet}}
\put(1727,1696){\makebox(0,0)[r]{$g\rightarrow \psi'$ Octet}}
\put(2008,51){\makebox(0,0){$z$}}
\put(100,1180){%
\makebox(0,0)[b]{\shortstack{$1/N dN/dz$}}%
}
\put(3417,151){\makebox(0,0){1}}
\put(2854,151){\makebox(0,0){0.8}}
\put(2290,151){\makebox(0,0){0.6}}
\put(1727,151){\makebox(0,0){0.4}}
\put(1163,151){\makebox(0,0){0.2}}
\put(600,151){\makebox(0,0){0}}
\put(540,1903){\makebox(0,0)[r]{4}}
\put(540,1490){\makebox(0,0)[r]{3}}
\put(540,1077){\makebox(0,0)[r]{2}}
\put(540,664){\makebox(0,0)[r]{1}}
\put(540,251){\makebox(0,0)[r]{0}}
\end{picture}

\vskip -7mm
\caption[dummy]{{\it The $z$-distribution for $20<p_{\tr\psi'}<25$ GeV as
    generated by {\sc Ariadne} for different production mechanisms.}}
\label{fig:z1}
\end{figure}
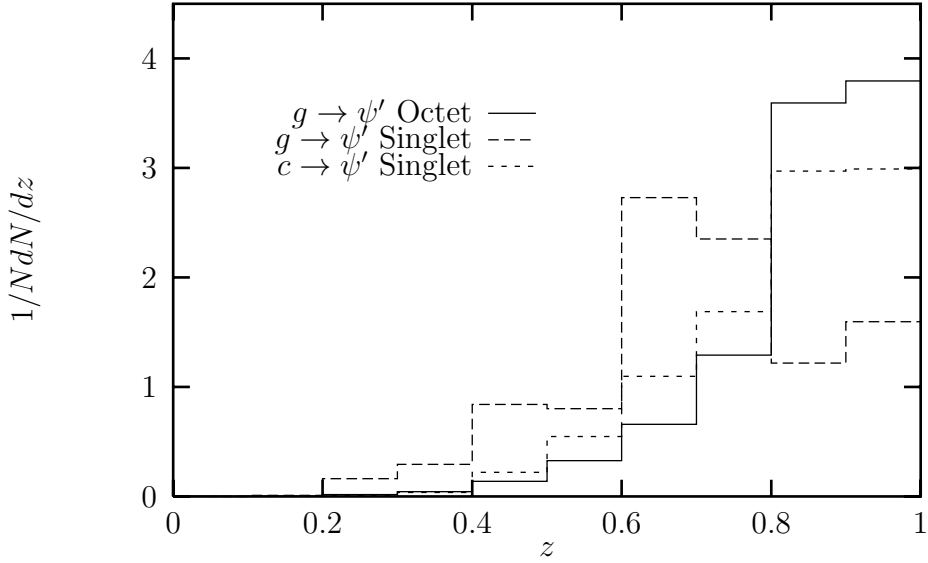

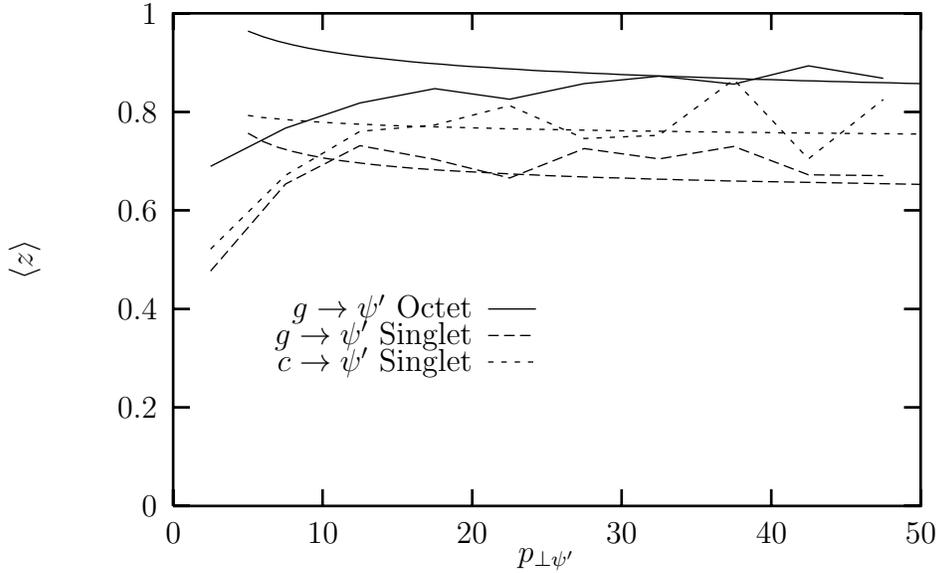
\begin{figure}

\setlength{\unitlength}{0.1bp}
\begin{picture}(3600,2160)(0,-100)
\put(1727,794){\makebox(0,0)[r]{$c\rightarrow \psi'$ Singlet}}
\put(1727,894){\makebox(0,0)[r]{$g\rightarrow \psi'$ Singlet}}
\put(1727,994){\makebox(0,0)[r]{$g\rightarrow \psi'$ Octet}}
\put(2008,51){\makebox(0,0){$p_{\perp\psi'}$}}
\put(100,1180){%
\makebox(0,0)[b]{\shortstack{$\langle z\rangle$}}%
}
\put(3417,151){\makebox(0,0){50}}
\put(2854,151){\makebox(0,0){40}}
\put(2290,151){\makebox(0,0){30}}
\put(1727,151){\makebox(0,0){20}}
\put(1163,151){\makebox(0,0){10}}
\put(600,151){\makebox(0,0){0}}
\put(540,2109){\makebox(0,0)[r]{1}}
\put(540,1737){\makebox(0,0)[r]{0.8}}
\put(540,1366){\makebox(0,0)[r]{0.6}}
\put(540,994){\makebox(0,0)[r]{0.4}}
\put(540,623){\makebox(0,0)[r]{0.2}}
\put(540,251){\makebox(0,0)[r]{0}}
\end{picture}

\vskip -7mm
\caption[dummy]{{\it The average $z$ as a function of $p_{\tr\psi'}$
    for different production mechanisms. The smooth lines are from the
    analytical approximation in the appendix and the (due to low
    statistics) broken lines were generated with {\sc Ariadne}.}}
\label{fig:z4}
\end{figure}

In fig.~\ref{fig:z1} we show the generated $z$-spectrum for
$20<p_{\tr\psi'}<25$ GeV, where $z$ is now $p_{\tr\psi'}/\hat{p}_\tr$
and $\hat{p}_\tr$ is the transverse momentum in the hard subprocess.
It is clear that the $z$--spectrum is much harder for the octet mechanism
as expected. This is also seen in fig.~\ref{fig:z4}, where the average
$z$ is shown as a function of $p_{\tr\psi'}$. Here we also find a good
agreement with the analytical approximation at large $p_{\tr\psi'}$.

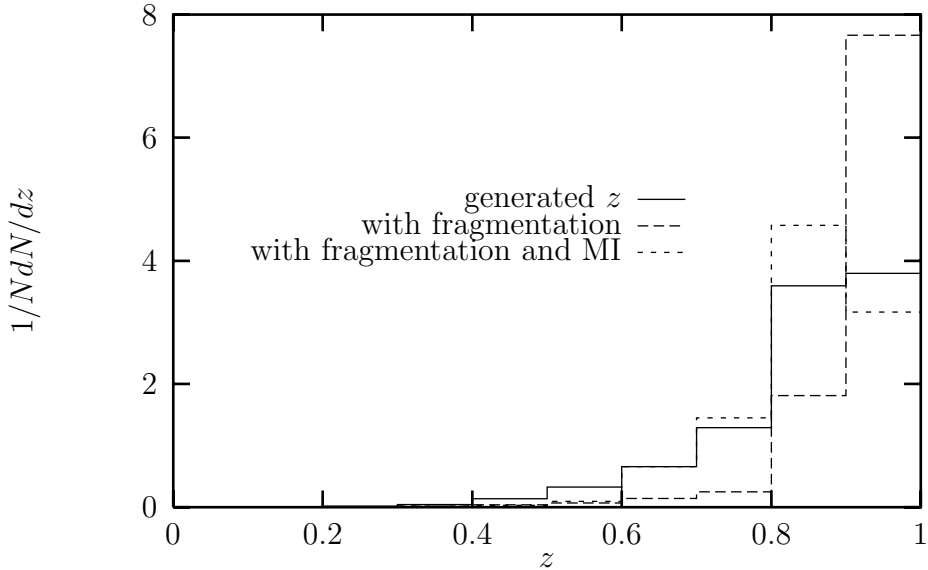
\begin{figure}

\setlength{\unitlength}{0.1bp}
\begin{picture}(3600,2160)(0,-100)
\put(2290,1212){\makebox(0,0)[r]{with fragmentation and MI}}
\put(2290,1312){\makebox(0,0)[r]{with fragmentation}}
\put(2290,1412){\makebox(0,0)[r]{generated $z$}}
\put(2008,51){\makebox(0,0){$z$}}
\put(100,1180){%
\makebox(0,0)[b]{\shortstack{$1/N dN/dz$}}%
}
\put(3417,151){\makebox(0,0){1}}
\put(2854,151){\makebox(0,0){0.8}}
\put(2290,151){\makebox(0,0){0.6}}
\put(1727,151){\makebox(0,0){0.4}}
\put(1163,151){\makebox(0,0){0.2}}
\put(600,151){\makebox(0,0){0}}
\put(540,2109){\makebox(0,0)[r]{8}}
\put(540,1645){\makebox(0,0)[r]{6}}
\put(540,1180){\makebox(0,0)[r]{4}}
\put(540,716){\makebox(0,0)[r]{2}}
\put(540,251){\makebox(0,0)[r]{0}}
\end{picture}

\vskip -7mm
\caption[dummy]{{\it The $z$ distribution for $20<p_{\tr\psi'}<25$ GeV
    as generated by {\sc Ariadne} using the color octet mechanism.
    Full line is the generated spectrum, long-dashed line is after
    fragmentation using a cone radius of 1 to define the associated
    jet, and short-dashed is the same with the addition of the
    underlying event according to the multiple scattering model in
    {\sc Pythia}.}}
\label{fig:z2}
\end{figure}

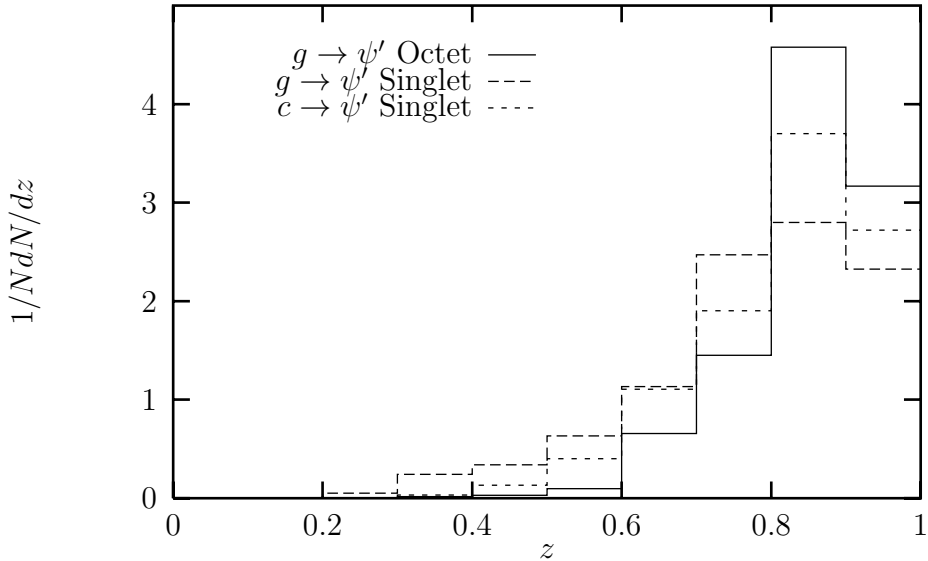
\begin{figure}

\setlength{\unitlength}{0.1bp}
\begin{picture}(3600,2160)(0,-100)
\put(1727,1723){\makebox(0,0)[r]{$c\rightarrow \psi'$ Singlet}}
\put(1727,1823){\makebox(0,0)[r]{$g\rightarrow \psi'$ Singlet}}
\put(1727,1923){\makebox(0,0)[r]{$g\rightarrow \psi'$ Octet}}
\put(2008,51){\makebox(0,0){$z$}}
\put(100,1180){%
\makebox(0,0)[b]{\shortstack{$1/N dN/dz$}}%
}
\put(3417,151){\makebox(0,0){1}}
\put(2854,151){\makebox(0,0){0.8}}
\put(2290,151){\makebox(0,0){0.6}}
\put(1727,151){\makebox(0,0){0.4}}
\put(1163,151){\makebox(0,0){0.2}}
\put(600,151){\makebox(0,0){0}}
\put(540,1737){\makebox(0,0)[r]{4}}
\put(540,1366){\makebox(0,0)[r]{3}}
\put(540,994){\makebox(0,0)[r]{2}}
\put(540,623){\makebox(0,0)[r]{1}}
\put(540,251){\makebox(0,0)[r]{0}}
\end{picture}

\vskip -7mm
\caption[dummy]{{\it The $z$ distribution for $20<p_{\tr\psi'}<25$ GeV as
    generated with {\sc Ariadne} with different production mechanisms,
    after hadronization and inclusion of multiple interactions.}}
\label{fig:z3}
\end{figure}

With the availability of $\psi'$ production in an event generator we
can now also study hadronization effects. To study eg.\ the
$z$-spectrum experimentally one could sum up the transverse energy
$E_{\tr\subjet}$ in a cone in the pseudorapidity--azimuth
plane around the detected $\psi'$, defining $z_{\subjet} =
p_{\tr\psi'}/E_{\tr\subjet}$. In fig.~\ref{fig:z2} we compare
the $z$-spectrum for such a procedure with the generated spectrum in
fig.~\ref{fig:z1} for the octet mechanism. We note that just adding
hadronization, a smearing of the jet is introduced and some of the jet
energy falls outside the cone, which makes the spectrum considerably
harder. Adding also multiple interactions to model the
underlying event\cite{TSmult87} in {\sc Pythia} makes it softer again. In
fig.~\ref{fig:z3} we see that the differences in hardness between the
octet and singlet mechanisms remains after all hadronization effects
although the shapes are somewhat distorted.

\section{Conclusions}
\label{sec:conclusions}

We have implemented different quarkonium production mechanisms in the
{\sc Ariadne} event generator. Besides giving a more accurate
treatment of the initial-- and final--state parton evolution than recent
analytical approximations, such a generator is an essential tool for
understanding experimental corrections and hadronization effects in
case one wants to study non-inclusive aspects of high $p_{\tr}$ onium
production at e.g.\ the Tevatron.

In interpreting the results it is important to keep in mind the status
of the implementations of the different production mechanisms:

\begin{itemize}
\item
The implementation of the color singlet $c \rightarrow \psi' + c$
fragmentation is unproblematic and trustworthy.

\item
The color singlet $g \rightarrow \psi' + gg$ fragmentation has been
implemented as $g \rightarrow \psi' + g$. The error introduced by
this simplification is believed to be small, and in the future the
full $g \rightarrow \psi' + gg$ splitting will be implemented.

\item
The color octet
$g\rightarrow c\bar{c} (^3S_1^{(8)}) \rightarrow \psi' \pm$ soft gluons,
has been implemented as $g\rightarrow \psi' +$ one soft gluon, and the
effects of the physical ${\cal O}(v^2)$ width of the fragmentation
function has not been investigated.
In the future the sensitivity of the results to the modelling of the
soft interactions will be studied.
\end{itemize}

The production mechanisms we have implemented here are, after a simple
change of normalization relevant also for $\psi$ production.
In this case it is necessary, however, to take into account also
contributions from P-wave states decaying to $\psi$.
It is therefore important to implement also P-wave production
mechanisms.

There are also other experiments where the fragmentation production of
quark\-onium states is important.  One interesting example is $Z^0$
decays at LEP.  Here the situation is quite different as compared to
large $\ptr$ production.  There are more quarks than gluons available
for fragmentation, and there is no trigger bias enhancement.
Consequently other production mechanisms might be dominant and
measurements would be more sensitive to the low $z$ shape of
fragmentation functions.

We have shown that it might be possible to measure the $z$-distribution
of the fragmentation function at the Tevatron and from that learn
about the production mechanism.

This result is of course dependent on our very crude approximation of
the experimental environment. But with the implementation presented
here, the determination of optimal choices of jet definitions and
experimental cuts is clearly facilitated and it does not seem unlikely
that the $z$ spectrum, if measured at the Tevatron, could be used to
better understand the quarkonium production mechanism.

{\bf Acknowledgement.}
We are greatful for discussions with Paul Hoyer, Michelangelo Mangano and
Mikko V\"anttinen.

\newpage
\section*{Appendix}
Here we present a very simple calculation that exhibits the gross
features of charmonium production at large $\ptr$.

We start out from a rapidity integrated version of eq.~(\ref{sigmap})
\beq {d\sigma_{\psi'}(\ptr) \over d\ptr} = \sum_i \int_0^1 dz
{d\sigma_i \over d{p_i}_T}(\ptr/z,\mu) D_{g \rightarrow
  {\psi'}}(z,\mu).
\label{sigma}
\eeq
We choose the factorization scale as $\mu=\ptr/z$, and note that
large logs $\log(4m_c^2/\mu^2)$ can be resummed by AP evolution
\cite{AP} of the fragmentation functions.
We calculate the parton production cross sections as
convolutions of hard subprocess cross sections ($\propto \ptr^{-4}$)
and parton distributions, and find that
the parton production cross sections fall roughly as
$\ptr^{-5}$.
Consequently we  take $d\sigma_i / d{p_i}_T \approx c_i / {{p_i}_T}^5 $
in eq.~(\ref{sigma}) and find
\beqa
  {d\sigma_{\psi'} \over d\ptr}
  & \approx &
     \sum_i
     \int_0^1 dz
    {c_i \over {\ptr}^5 } z^5
    D_{i \rightarrow {\psi'}}(z,{p_i}_T/z)
\approx
\sum_i
{ c_i \over \ptr^5 }
D_{i \to\psi'}^{(6)}(\ptr)  .
\label{sigmaMom}
\eeqa
Here $D_{i \to\psi'}^{(6)}(\mu) \equiv
\int_0^1 dz z^5 D_{i \rightarrow {\psi'}}(z,\mu)$ is the sixth moment of the
fragmentation function.
In eq.~(\ref{sigmaMom}) we have neglected the evolution of the fragmentation
functions between $\mu=\ptr$ and $\ptr/z$, implicitly assuming that
the dominant contributions to the integral in eq.~(\ref{sigmaMom})
come from large values of $z$.

We find $c_g \approx 7 \cdot 10^9 \gev^4 {\rm nb}$, and
$c_c \approx 4 \cdot 10^7 \gev^4 {\rm nb}$, to LO in $\as$.
The gluon production cross section is two orders of
magnitude larger than the quark production cross section.
Within the color singlet model to LO in $\as$, the c quark fragmentation
function is, however, much larger than the gluon fragmentation function.
We keep therefore both gluon and c quark fragmentation in
eq.~(\ref{sigmaMom}).

In \cite{BDFMhighptSingletTheory} mixing effects were taken into account
by including NLO corrections to the hard scattering charm production
cross section.
Here as well as in our QCD shower generator implementation, we include
mixing explicitly and use the LO charm production cross section to avoid
double counting.

For moments ($n>1$) the AP evolution equations can be solved
analytically. The solution has the form:
\begin{eqnarray}
  D_{g \to\psi'}^{(n)}(\mu)
  & = &
  \left( \as(\mu) \over \as(\mu_0) \right)^{d_g^{(n)}}
  \left( (1-\epsilon^{(n)}) D_{g \to\psi'}^{(n)}(\mu_0)
    -  \delta^{(n)}   D_{c \to\psi'}^{(n)}(\mu_0) \right)
  \nonumber
  \\ & &
  +
  \left( \as(\mu) \over \as(\mu_0) \right)^{d_c^{(n)}}
  \left( \epsilon^{(n)} D_{g \to\psi'}^{(n)}(\mu_0)
    + \delta^{(n)}   D_{c \to\psi'}^{(n)}(\mu_0) \right)
  \label{Dg6solution}
  \\
  D_{c \to\psi'}^{(n)}(\mu)
  & = &
  \left( \as(\mu) \over \as(\mu_0) \right)^{d_c^{(n)}}
  \left( (1-\epsilon^{(n)}) D_{c \to\psi'}^{(n)}(\mu_0)
    +  \rho^{(n)}     D_{g \to\psi'}^{(n)}(\mu_0) \right)
  \label{Dc6solution}
  \nonumber \\ & & + \left( \as(\mu) \over \as(\mu_0)
  \right)^{d_g^{(n)}} \left( \epsilon^{(n)} D_{c \to\psi'}^{(n)}(\mu_0)
    - \rho^{(n)} D_{g \to\psi'}^{(n)}(\mu_0) \right) ,
\end{eqnarray}
with $\rho^{(n)} \delta^{(n)} = \epsilon^{(n)} (1-\epsilon^{(n)})$.
For the moments $n$ relevant here, the mixing parameters $\rho^{(n)}$,
$\delta^{(n)}$ and $\epsilon^{(n)}$ are all small, and we may put
$\rho^{(n)}$ and $\epsilon^{(n)}$ to zero.  Within the color singlet
model the $\delta^{(n)} D_{c \to\psi}^{(n)}$ terms in
eq.~(\ref{Dg6solution}) are important however since as noted above the
c quark fragmentation function is much larger than the gluon
fragmentation function.
For simplicity we choose the initial scales for charm and gluon fragmentation
and the mixing onset scale all to $2m_c$ and ignore flavor thresholds
(refinements are simple).
Combining eq.~(\ref{sigmaMom}),(\ref{Dg6solution}) and (\ref{Dc6solution})
we find
\begin{eqnarray}
  {d\sigma_{\psi'} \over d\ptr}
  & \approx &
  { c_g \over \ptr^5}
  \left( \as(\ptr) \over \as(2m_c) \right)^{d_g^{(6)}}
  \left( D_{g \to\psi'}^{(6)}(2m_c) -
    \delta^{(6)} D_{c \to\psi'}^{(6)}(2m_c) \right)+
  \nonumber
  \\ & &
  { c_c + \delta^{(6)} c_g \over \ptr^5}
  \left( \as(\ptr) \over \as(2m_c) \right)^{d_c^{(6)}}
  D_{c \to\psi'}^{(6)}(2m_c) .
  \label{sigmafinal}
\end{eqnarray}
According to the hypotheses by Braaten \etal \cite{BraatenFleming}
the color octet
mechanism gluon fragmentation completely dominates $\psi'$ production.
We may then of course ignore charm fragmentation all together in
eq.~(\ref{sigmafinal}).

Similarly we can calculate the mean value of $z$ in the convolution in
eq.~(\ref{sigma}).  Superficially identifying $z$ with
$p_{\tr \psi'}/p_{\tr\subjet}$, this gives us a hint
of the possibilities to use the form of fragmentation functions to
distinguish between different production mechanisms.  More
importantly, it can be used to check our QCD cascade implementation of
quarkonium production.

For convenience we give numerical values for the moment evolution
parameters $\delta^{(n)}$, $d_g^{(n)}$, $d_c^{(n)}$ for 4 (5) quark
flavors, as well as the values we have used for the
moments of the color singlet fragmentation functions
$D_{c\rightarrow {\psi'}}^{(n)}$ and $D_{g\rightarrow {\psi'}}^{(n)}$:

\begin{minipage}[b]{\textwidth}
  \renewcommand{\arraystretch}{1.2}
  \begin{eqnarray*}
    \begin{array}{|l|c|c|c|}
      \hline
      n  & 5 & 6 & 7
      \\ \hline \hline
      \delta^{(n)} & 0.030 \; (0.028) & 0.023 \; (0.022) & 0.018 \; (0.017)
      \\ \hline
      d_g^{(n)} & 2.18 \; (2.46) & 2.46 \; (2.76)  & 2.68 \; (3.00)
      \\ \hline
      d_c^{(n)} & 0.97 \; (1.05) & 1.08 \; (1.17) & 1.17 \; (1.27)
      \\ \hline
      D_{c\rightarrow {\psi'}}^{(n)}(2m_c)
         & 1.2 \cdot 10^{-5} & 9.7 \cdot 10^{-6} & 7.7 \cdot 10^{-6}
      \\ \hline
      D_{g\rightarrow {\psi'}}^{(n)}(2m_c)
         & 1.8 \cdot 10^{-7} & 1.3 \cdot 10^{-7} & 1.1 \cdot 10^{-7}
      \\ \hline
    \end{array}
    & & \nonumber
  \end{eqnarray*}
  \renewcommand{\arraystretch}{1}
\end{minipage}
For the moments of the color octet gluon fragmentation function we have
used the value
$D_{g\rightarrow {\psi'}}^{(n)}(2m_c) \approx 4.2 \cdot 10^{-5}$,
independent of the moment $n$.

\end{document}